\documentclass[prb,superscriptaddress,preprint,floatfix]{revtex4-1}
\usepackage{amsmath}
\usepackage{amsfonts}
\usepackage{amssymb}
\usepackage{graphicx}
\usepackage{subfig}
\newcommand{\etal}{\emph{et al.}}
\newcommand{\vgr}{\emph{v.gr. }}
\newcommand{\eg}{\emph{e.g. }}
\newcommand{\ie}{\emph{i.e. }}

\newcommand{\hn}{$h\nu$}

\newcommand{\mum}{\: \mu m}
\newcommand{\film}{As$_2$S$_3$}
\newcommand{\vs}{\emph{versus}}
\newcommand{\e}{\varepsilon}
\newcommand{\er}{\e_r}
\newcommand{\ur}{\mu_r}
\newcommand{\kphotonics}{Photonics Laboratory, King Abdullah University of Science and Technology \emph{(KAUST)}, Thuwal 21534, Saudi Arabia}
\newcommand{\kfupm}{King Fahd University of Petroleum and Minerals \emph{(KFUPM)}, Dhahran 31261, Saudi Arabia}
\newcommand{\anic}{Advanced Nanofabrication and Imaging Core-Lab, King Abdullah University of Science and Technology \emph{(KAUST)}, Thuwal 21534, Saudi Arabia}

\begin{document}
\title{Electron irradiation induced reduction of the permittivity in chalcogenide glass (As$_2$S$_3$) thin film}
\author{\firstname{Dami\'an P.} \surname{San-Rom\'an-Alerigi}}
\affiliation{\kphotonics}
\email[Corresponding author: ]{damian.sanroman@kaust.edu.sa}
\author{\firstname{Dalaver H.} \surname{Anjum}}
\affiliation{\anic}
\author{\firstname{Yaping} \surname{Zhang}}
\affiliation{\kphotonics}
\author{\firstname{Xiaoming} \surname{Yang}}
\affiliation{\anic}
\author{\firstname{Ahmed} \surname{Benslimane}}
\affiliation{\kphotonics}
\author{\firstname{Tien K.} \surname{Ng}}
\affiliation{\kphotonics}
\author{\firstname{Mohamed N.} \surname{Hedhili}}
\affiliation{\anic}
\author{\firstname{Mohammad} \surname{Alsunaidi}}
\affiliation{\kphotonics}
\affiliation{\kfupm}
\author{\firstname{Boon S.} \surname{Ooi}}
\affiliation{\kphotonics}

\begin{abstract}
\emph{In this paper we investigate the effect of electron beam irradiation on the dielectric properties of \film~Chalcogenide glass. By means of low-loss Electron Energy Loss Spectroscopy, we derive the permittivity function, its  dispersive relation, and calculate the refractive index and absorption coefficients under the constant permeability approximation. The measured and calculated results show a heretofore unseen phenomenon: a reduction in the permittivity of $\geq40\%$.  Consequently a reduction of  the refractive index of $20\%$, hence suggesting a conspicuous change in the optical properties of the material under irradiation with a $300\:\mathrm{keV}$ electron beam. The plausible physical phenomena leading to these observations are discussed in terms of the homopolar and heteropolar bond dynamics under high energy absorption. The  reported phenomena, exhibited by \film-thin film, can be crucial for the development of photonics integrated circuits using electron beam irradiation method. \emph{Published at Journal of Applied Physics {\bf 113}, 044116, 2013.  DOI: 10.1063/1.4789602}}
\end{abstract}
\maketitle

\section{Introduction\label{sec.intro}}

Over the last  decades, chalcogenide glasses have been a subject of great interest  due to the myriad photon-induced phenomena they exhibit, captivating the imagination of scientists and engineers to prompt  countless photonic applications ranging from biology\cite{Yang2010,Eggleton2011} and telecommunications\cite{Pelusi2010a,Eggleton2010}, all-optical chips\cite{Eggleton2010,Eggleton2011}, single photon sources\cite{Xiong2010,Bendana2011} to holography\cite{Juodkazis2004}. Furthermore, they have been worthy of great attention from fundamental science, owing to the seemingly oxymoronic nature of the physical effects observed when the material is illuminated by photons within its band-gap energy range ($E_g\:\sim\:2.4\mathrm{eV}$) , \eg giant photo-expansion\cite{Chen1970} \vs~ photo-contraction\cite{Kozak2009a}, photo-liquidity\cite{Fritzsche1996a,Tanaka2005} \vs~ photo-crystallization\cite{Feinleib1971,Tanaka2009}, photo-darkening\cite{Istvan2007,DeNeufville1974} \vs~ photo-bleaching\cite{Istvan2007}, and, more frequently, photo-refraction\cite{DeNeufville1974,Gopal1982,Kurtz2009,VKTikhomirovandSRElliott1995}, to cite a few. Studies from  fundamental physics and material science have led to various models attempting to  explain the mechanisms behind the exciting structural reconfiguration capabilities and phenomena observed in chalcogenide glasses upon energy absorption. That being said, the physical processes, in addition to many of the phenomena reported,  remain both a matter of debate and a hot topic in fundamental and applied research.\\

The principal experiments, observations and models in chalcogenide glass deal with phenomena triggered by light irradiation. Photon induced phenomena in chalcogenide glass comprises: \emph{photon-induced} -refractive index modification, -liquidity, -dichroism, -anisotropy, -crystallization, -darkening, -bleaching, and -vitrification \cite{Chen1970, Kozak2009a,Fritzsche1996a,Tanaka2005,Feinleib1971,Tanaka2009,Istvan2007,DeNeufville1974, Gopal1982,Kurtz2009,VKTikhomirovandSRElliott1995,Submitted2010}, in addition to other innate effects present in highly non-linear materials, \vgr second harmonic generation \cite{Ta'eed2006, Lyubin2006} for example. To understand  the nature of these effects, and the electronic and atomic processes involved, it is essential to investigate the charge carrier transfer, energy spectrum, and the mechanisms by which  radiation (electron or photon) interacts with the material modifying its chemical and atomic structure \cite{Kozak2009a,Tanaka2005,Tanaka2009,Anderson1975,Fritzsche1998,Simdyankin2004,Andriesh2004,Singh2007}.\\

Electron irradiation induced refractive index modification has also previously been observed. Suhara \etal \cite{Suhara1975},  found a maximum change of $\Delta n/n = +3.6\%$, with no significant modification in the  thickness of the film. On the other hand, Normand \etal\cite{Nordman1996a,Nordman2001a,Nordman2001} and Tanaka \etal\cite{Tanaka1997a}, independently observed a $\sim 3\%$ increment in the refractive index of chalcogenide glass;  simultaneously the formation of trenches and mounds, $180\:\mathrm{nm}$ and $110\:\mathrm{nm}$ respectively, in chalcogenide films of $5 -11\mum$ thick, were observed. They posited that the morphological and optical alteration derives from the structural reorganization and re-bonding of the homopolar and heteropolar bonds, in addition to the electrostatic effects arising from the charge density variation \cite{Nordman1996a,Nordman2001a,Nordman2001,Tanaka1997a}.\\

Whilst many of these effects appeal to our research interest, we are particularly intrigued  by the alteration in the optical properties of chalcogenide glass thin films, and its possible  applications. However, in most of the previous studies only the refractive index has been characterized, with little information available on how the electromagnetic components, \ie \emph{permittivity} and \emph{permeability}, reshape under energy absorption. In this paper we present the results of our investigation on the permittivity dynamics of chalcogenide glass  \film~thin film under energy absorption by means of low-loss electron energy loss spectroscopy (EELS). We believe that understanding the electromagnetic characteristics, and their dynamics, particular to  this material are key to enabling and extending many of the devices proposed by transformation optics and achieving light control beyond the scope of [meta]materials.\\

To this effect, we have organized the paper as follows; Section \ref{sec.intro} presented a review; section \ref{sec.methods} introduces the theoretical and experimental methodology needed to explain the experimental results;  section \ref{sec.analysis} presents the results and analysis, followed by the deduction of the refractive index under the constant permeability approximation, while discussing its implications; finally, section \ref{sec.fin}, we present our conclusions.\\

\section{Methods\label{sec.methods}}

It is worth to discern the difference, and connection, between the refractive index and the electromagnetic properties of materials. The former is the ratio between the propagation speed of waves inside a material to that of vacuum; whereas the latter, permittivity ($\e$) and permeability ($\mu$), are the physical representation of a material's intrinsic electric and magnetic properties; which arise from the atomic, electronic and molecular configuration and interactions with the electric and magnetic fields, ultimately shaping their propagation dynamics.  Accordingly, the refractive index and the electromagnetic properties of materials are entwined by the widespread relation: $n=\sqrt{\er\ur}$, where the subscript $r$ refers to the relative permittivity or permeability, \ie $\er=\e/\e_0$ and $\ur=\mu/\mu_0$.\\

Permeability and permittivity are, then, key to manipulate light propagation in materials, and hence the importance of characterizing them, and understanding how can they be modified.

\subsection{Electron Energy Loss Spectroscopy}

As discussed earlier, \film~chalcogenide glass, exhibits electron- and photon- irradiation induced modification of its optical properties. In our study we use low-loss EELS\cite{Perrin1974, Egerton1996} to characterize the real and imaginary parts of the permittivity of \film.\\

When high energy electrons, $~ 300$keV impinge on a sufficiently thin material, thickness $\leq 300nm$, a vast majority of the electrons will pass through the sample without being perturbed, and a small fraction of them will undergo inelastic scattering loosing energy to the sample, typically in the order of $10^1-10^2\: \mathrm{eV}$. It is important to note, however, that some of the electrons will also be elastically scattered and will  transfer some energy to the system; yet, this energy is not enough to perturb the overall atomic arrangement of the material, even on a head-on collision with the atom the elastic scattered electron energy loss will be in the range of $~1\:\mathrm{eV}$, which is not sufficient to displace the atoms and therefore change the molecular arrangement. The inelastic scattering, nevertheless, yields enough energy to the system to sustain bond-breaking and alterations in the electron density of the material. The latter electron loss can be experimentally measured via an electron spectrometer.\\ 

The inelastic scattering process is characterized by the electron scattering angle, $\theta$, or momenta, $\vec q$, and the electron energy loss, $E_s$, due to their interactions with the material, which includes phonon excitation, inter- and intra- band transition, plasmons excitation, inner shell ionization, and Cherenkov radiation. These interactions can be summarized by the dielectric response function, $\e(\vec q, \omega)$; which describes the interaction of photons and electrons with the material \cite{Egerton1996}.\\

Admittedly, the interaction of photons and electrons with those in the solid is different; photons displace the electrons in the material in a direction perpendicular to their direction of propagation, with the electron density remaining unchanged. Therefore, we define the \emph{optical permittivity} as a transverse property of a medium; whereas electrons interacting with the material produce a longitudinal displacement of the electrons, and change the electron density. Moreover, the energy transfer mechanisms for photons differs from that of electrons; in the former the energy transfer is mostly mediated by inter- and intra- band transitions (valence-to-conduction band), \eg (i) bandgap absorption, (ii) defects, (iii) free carriers, among other transitions. In the scenario of electron irradiation, energy transfer takes place by means of inelastic scattering, and to a lesser degree by elastic scattering. Despite the fundamental differences, however, it is possible to establish a correlation between the permittivity measured by electrons or photons if we constrain to small energy losses, which translates small scattering angles for electrons. Under this condition $\e(\vec q, \omega)$ varies insignificantly with $\vec q$, and hence $\e(\vec q, \omega)\approx \e(\vec q = 0, \omega=E/\hbar)$, the latter being the optical permittivity \cite{Egerton1996, Ritchie1957, Nozieres1959}.\\

Whence stems the versatility of low-loss EELS over other methods to measure the permittivity of thin film samples constraint to small probing areas. Specifically, if scanning transmission microscopy (STEM) is used, the volume of study can be reduced to nano-metric scale, allowing us to study the material and its response locally. As discussed in the preceding paragraph in a typical EELS experiment low angles of collections are used, and thus most inelastic collisions result in energy losses to the incoming electrons of less than $100\:\mathrm{eV}$. This energy loss range in turn deciphers information about the permittivity within an energy range of interest to optics and photonics, \ie $\leq 10\:\mathrm{eV}$. The latter is achieved by post-processing the energy loss spectra by means of Kramers-Kronig relations  to evince the real and imaginary parts of the optical permittivity \cite{Egerton1996,Stoger-Pollach2008,Egerton2009,Verbeeck2009, GarciadeAbajo2010,Zhang2010b,Schafer2012}.\\

\subsection{Experimental arrangement}
  
In our experiment, high energy electrons ($~300\:{keV}$) impinge on an \film~thin film, $300\:\mathrm{nm} \pm\:5\:\mathrm{nm}$ thick, grown on a $2\:\mu m$ holey-Cu TEM grid by means of electron-beam evaporation, and the roughness, film thickness and stoichiometry are characterized by atomic force microscopy (AFM),  ellipsometry, energy dispersive X-ray (EDX), and X-ray photoemission spectroscopy (XPS), respectively.\\

To ensure the validity of the low energy loss limit approximation, we use a scanning transmission electron microscope (STEM), coupled to a low-loss EELS, and perform the experiment in the low momentum transfer relativistic approximation. We set the electron source to $300\:\mathrm{keV}$, with a collection angle of $6\:\mathrm{mrad}$, and  incident semi-angle of $4.575\:\mathrm{mrad}$. Under these conditions the energy resolving power of the electron loss spectrometer is better than $0.1\:\mathrm{eV}$ in the range $0.8\:\mathrm{eV} - 50\:\mathrm{eV} $; electrons are accelerated to $77.561\%$ the speed of light, corresponding to a de Broglie wavelength of $1.9687\times 10^{-3}\:\mathrm{nm}$, which results in a minimum spatial resolution of $1.2048\times 10^{-1}\:\mathrm{nm} $. The scanning area for the STEM is $50 \times 50 \:\mathrm{nm^2}$ on the holey region of the film support to avoid  any contamination from the carbon support of the grid. The density of electrons per second incident on the sample is 175 $e^-/\mathrm{nm}^2 \mathrm{s}$. To achieve different electron dosages we control the irradiation time in steps of $100\:\mathrm{ms} \pm .05 \:\mathrm{ms}$ (see table \ref{table.dosage}). These steps are divided into two irradiation sequences, sequence $A$ and sequence $B$; which are separated by a \emph{relaxation} time defined as the period of time where zero electrons are incident on the film. This arrangement allows us to determine if after a given time of repose the material returns to its initial state, \ie if it exhibits memory or a hysteretic behaviour.\\ 
\begin{table}[h!]
	\centering
	\begin{tabular}{l | c c c c | c | c c}
	\hline
	~ & \multicolumn{4}{c|}{\emph{Sequence A}} & \multicolumn{1}{c|}{\emph{Relaxation}} & \multicolumn{2}{c}{\emph{Sequence B}}\\
	Irradiation steps (\emph{step}) & 1 & 2 & 3 & \multicolumn{1}{c|}{4} & \multicolumn{1}{c|}{pause} & 5 & 6 \\ \hline
	Irradiation exposure time $t^*\: (\mathrm{s})$ & 0.1 & 0.2 & 0.3 & 0.5 & 60 & 0.8 & 1.0 \\
	Accumulated irradiation time $t\: (\mathrm{s})$ & 0.1 & 0.3 & 0.6 & 1.1 & - & 0.8 + 1.1 & 1.8 + 1.1\\
	\hline
	\end{tabular}
\caption{Irradiation sequence summary, including steps, exposure, and accumulated time. The symbol ``+'' is used to demark the fact that the material was exposed to irradiation sequence A and relaxation before the new measurement. \label{table.dosage}}
\end{table}\\

\section{Observations and analysis\label{sec.analysis}}

As discussed earlier in the text, we study the energy loss spectra of transmitted electrons in the low momentum transfer relativistic approximation, customarily known as low-loss EELS. The experiment results is a spectra-like graph detailing the density of electrons versus their output energy at different irradiation dosages.\\ 

\subsection{Results analysis}

\begin{figure}[hb]
	\centering
	\includegraphics[width=0.7\textwidth]{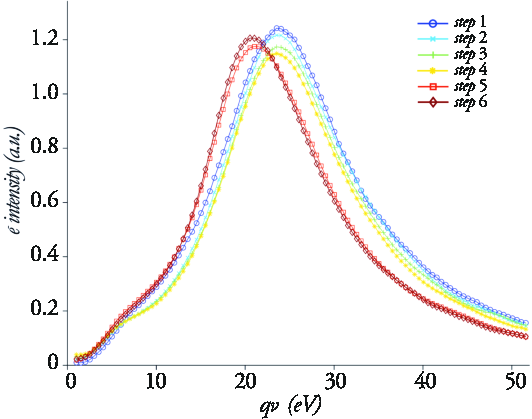}
	\caption{Low-loss Electron Energy Loss spectra, raw measurement results at different dosages as specified in table \ref{table.dosage}. The causal variable marks the energy lost by the electrons to the material through inelastic scattering, $q\nu$.\label{fig.lossraw}}
\end{figure}

In order to obtain the optical and electronic characteristics of the material, the loss spectra needs to be deconvoluted, and the zero loss subtracted: this is attained by the Fourier-Log Two Gaussian method \cite{Egerton1996} (see figure \ref{fig.lossraw}).\\

\begin{figure}[hb]
	\centering
	\subfloat[{~}]{\includegraphics[width=0.5\textwidth]{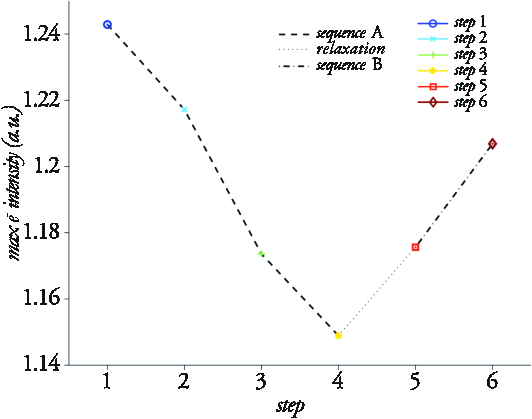}\label{fig.lossraw-sampled}}
	\subfloat[{~}]{\includegraphics[width=0.5\textwidth]{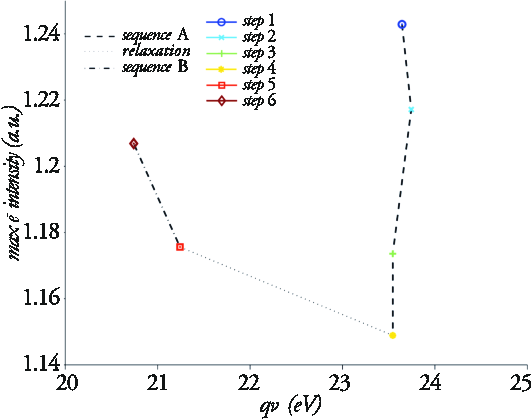}\label{fig.lossraw-dispersion}}
	\caption{Analysis of peak loss for steps $1$ through $6$, the behaviour of the peak in reference to the irradiation step is shown.\label{fig.lossraw-analysis}}
\end{figure}

A meticulous inspection of figure \ref{fig.lossraw} reveals that the peak of the loss spectra decreases as a function of electron dosage, as depicted in figure \ref{fig.lossraw-sampled}, after irradiation sequence A, and before relaxation, the peak value changes $8 \%$. It is also evident that the process has some degree of elasticity, where after relaxation some recovery occurs such that the electron intensity before and after relaxation suffers a $2\%$ return (see figure \ref{fig.lossraw-sampled}). However, this return goes hand in hand with a blue-shift of the peak dispersion of $\sim 3\:\mathrm{eV}$ (see figure \ref{fig.lossraw-dispersion}). Yet there is a striking effect, subsequent irradiation for long periods of time has a reverse effect, increasing the total electron intensity rather than decreasing it as was the case before relaxation in sequence A. As we will elucidate later this effect can be explained by the breaking of the homopolar bonds, and the continuous reconfiguration of the heteropolar bonds within the material. It is capital to recall that our experiment is carried in a vacuum environment ($10^{-8}\: \mathrm{Torr}$), therefore eliminating any chance of oxidation and contamination on the material surface.\\

\begin{figure}[hb]
	\centering
	\subfloat[{~}]{\includegraphics[width=.5\textwidth]{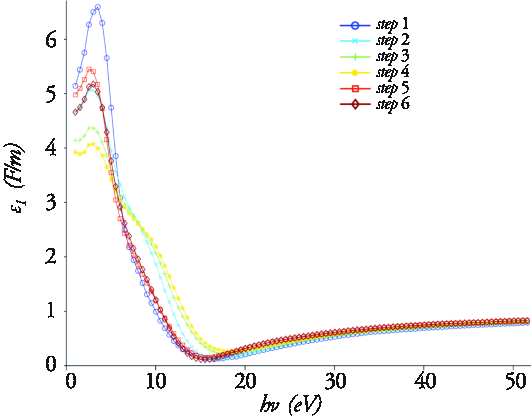}\label{fig.e1raw}}
	\subfloat[{~}]{\includegraphics[width=.5\textwidth]{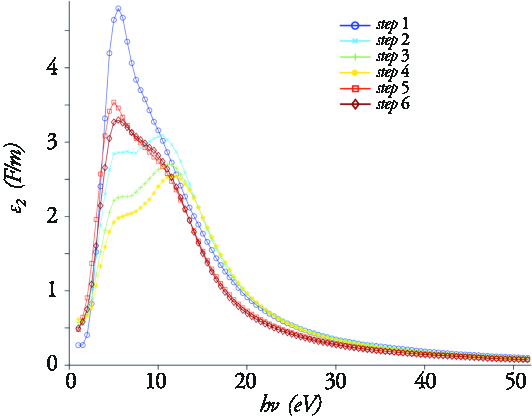}\label{fig.e2raw}}\\
	\subfloat[{~}]{\includegraphics[width=.5\textwidth]{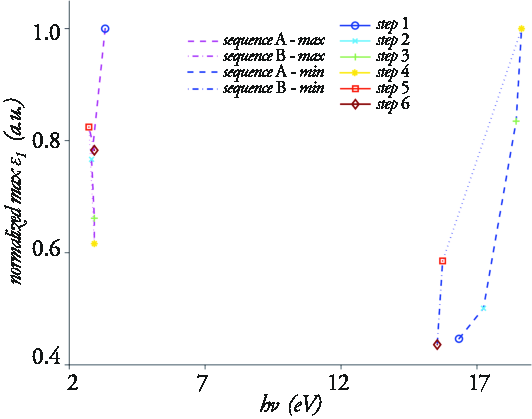}\label{fig.e1-dispersion}}
	\subfloat[{~}]{\includegraphics[width=.5\textwidth]{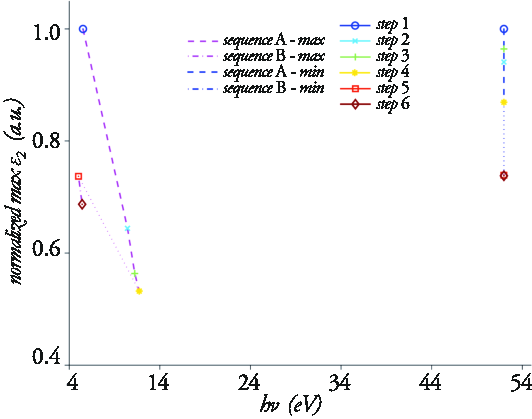}\label{fig.e2-dispersion}}
	\caption{(a) Real ($\e_1$) and (b) imaginary ($\e_2$) parts of the transversal permittivity function derived via Kramers-Kronig relations from the electron energy loss spectra. The peak trace, \ie dispersion relations, for (c) $\e_1$, and (d) $\e_2$, clearly show the semi hysteretic behaviour of the permittivity.\label{fig.eelsresults}}
\end{figure}

As discussed earlier, the electron energy loss is linked to the real and imaginary permittivity (see equation \ref{eq.sigmaepsilon}), and hence we can expect them to follow a somewhat similar pattern to that of the energy loss. \cite{Egerton1996}.\\ 

\begin{equation}\label{eq.sigmaepsilon}
	\sigma_{eels}=\Im(-1/\e)=\frac{\e_2}{\sqrt{\e_1^2+\e_2^2}}.
\end{equation}\\
\indent By virtue of the above equation and the Kramers-Kronig relations we obtain the \emph{imaginary} part of the permittivity function, $\e_2=\Im(\e)$, and the \emph{real} part, $\e_1=\Re(\e)$, here shown in figure \ref{fig.eelsresults}.\\

It is essential for the ongoing discussion to remark that in the curves of figure \ref{fig.eelsresults}, and subsequent, the causal variable, \hn, refers to the photon energy, for recall that these result from the Kramer-Kronig analysis of the electron energy loss spectra at the low momentum approximation. Therefore, they can be regarded as dispersion relations for the permittivity at different electron irradiation dosages.\\ 
 
As predicted at analyzing the electron energy loss spectra, the real and imaginary parts of the permittivity decrease as the electron dosage augments. Moreover, figures \ref{fig.e1-dispersion} and \ref{fig.e2-dispersion} confirm the semi hysteretic comportment  of the permittivity. Particularly, observe that the maxima of $\e_1$ reduces, and shifts following a semi hysteresis loop, as the total dosage rises; while its minima shifts to higher energies, although seldom changing its dispersion energy \hn. On the other hand, the maxima of $\e_2$ exhibits a similar decrement, as that portrayed by $\e_1$ as a function of dosage, with the difference that there is no evidence of a hysteretic behaviour, rather the maximum shifts to higher energies together with electron dosage.\\

\begin{figure}[hb]
\centering
\subfloat[$\e_1$ at range A]{\includegraphics[width=0.5\textwidth]{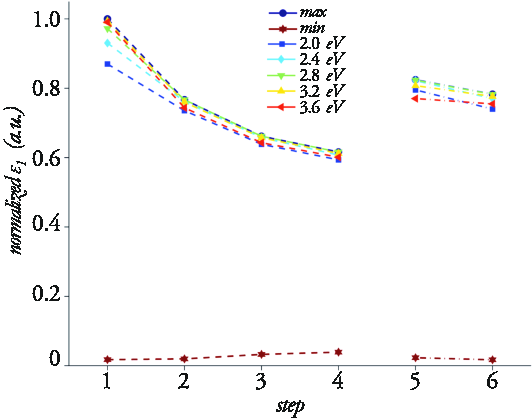}\label{fig.e1-bg}}
\subfloat[$\e_1$ at range B]{\includegraphics[width=0.5\textwidth]{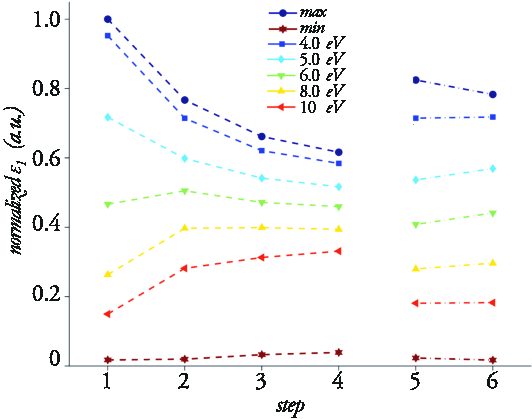}\label{fig.e1-uv}}\\
\subfloat[$\e_2$ at range A]{\includegraphics[width=0.5\textwidth]{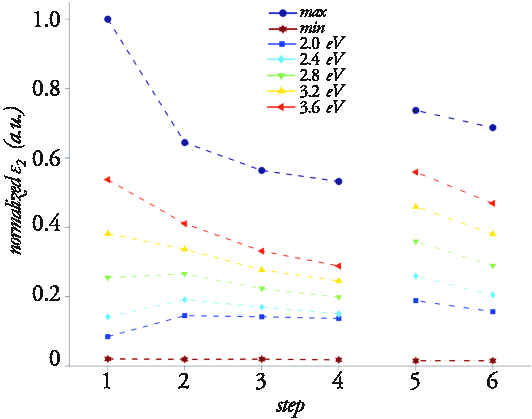}\label{fig.e2-bg}}
\subfloat[$\e_2$ at range B]{\includegraphics[width=0.5\textwidth]{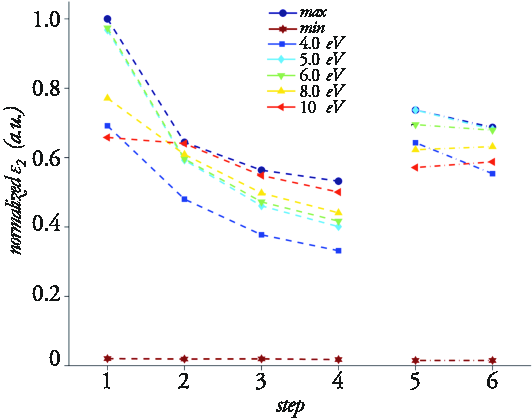}\label{fig.e2-uv}}
\caption{Selected samples of the permittivity function $\e$  (real and imaginary) in reference to the dispersion energy, \emph{range A} (a) and (c), and \emph{range B} (b) and (d), at different irradiation dosages. \label{fig.pngilon-extreme}}
\end{figure}

\begin{figure}[hb]
\centering
\subfloat[$\partial_{t^*}\e_1$]{\includegraphics[width=0.5\textwidth]{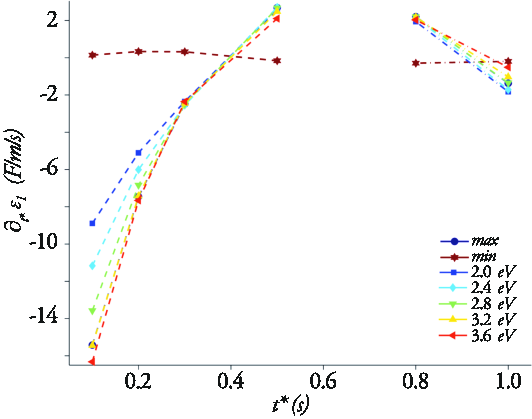}\label{fig.e1-bg-diff}}
\subfloat[$\partial_{t^*}\e_1$]{\includegraphics[width=0.5\textwidth]{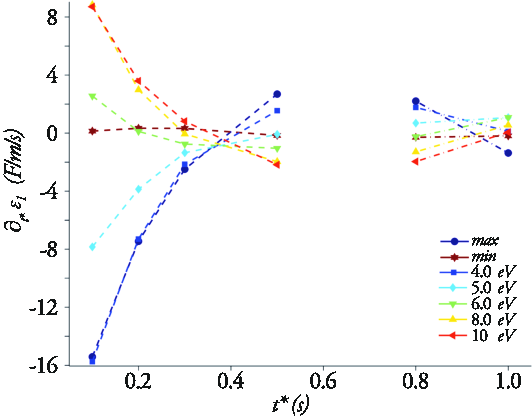}\label{fig.e1-uv-diff}} \\
\subfloat[$\partial_{t^*}\e_2$]{\includegraphics[width=0.5\textwidth]{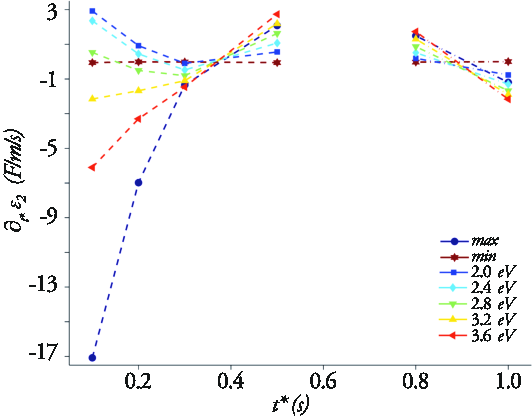}\label{fig.e2-bg-diff}}
\subfloat[$\partial_{t^*}\e_2$]{\includegraphics[width=0.5\textwidth]{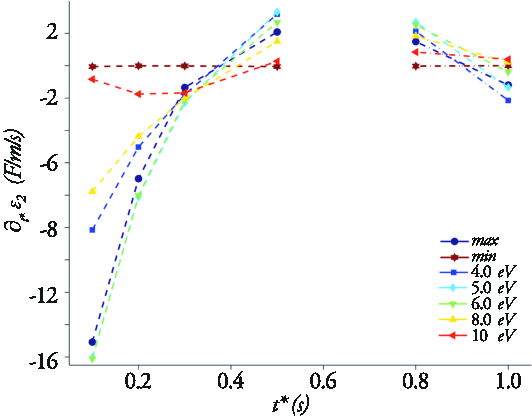}\label{fig.e2-uv-diff}}
\caption{Derivatives of the permittivity function with respect to the causal variable dosage, $t^*$. \label{fig.derivatives}}
\end{figure}

To acquire a deeper insight on the  permittivity of the material and its dynamics we sampled it by splitting the dispersion energy range  in two, \emph{range A} comprises energy values around the \emph{bandgap} of the material, \ie $2.0eV\leq h\nu \leq 3.6eV$; and \emph{range B} includes higher energies $4eV\leq h\nu \leq 10eV$.  At either set we compare the results to the global maximum and minimum, denoted by \emph{max} and \emph{min}, receptively.  Within \emph{range A} the sampling results,  figures \ref{fig.e1-bg} and \ref{fig.e2-bg}, clearly evince the hysteretic behaviour of the electron-induced permittivity change. At comparing the curves given by the maximum peaks of $\e_1$ and $\e_2$, before and after relaxation, it is patent that some degree of restoration occurs in the system which allows a recovery of up to $75\%$ of the initial values. Furthermore, the dynamics of the extreme values of either the real or imaginary permittivity echo an almost identical pattern, with an initial fast rate of change during the first irradiation steps $0\leq t^*\leq0.2\:\mathrm{ms}$, closely followed by a sustained decrement on its absolute value, \ie the pace at which the permittivity changes tends to zero for a sufficiently large irradiation time (see figures \ref{fig.derivatives}). As we will explain in section \ref{sec.explanation} this effect can be understood through the variation in the density and rearrangement of the homopolar and heteropolar bonds, which significantly modifies the response of the material to energy absorption.\\

As it can be appreciated at comparing the results between the two energy ranges, either for real or imaginary parts of the permittivity, shows significantly different comportments. The test values for the real permittivity in \emph{range A}  closely follow the pattern and magnitude modification of the global maximum, while the global minimum remains virtually unperturbed. In \emph{range B}, however, the selected values present different bearing to that observed previously. Specifically, before relaxation, samples at $h\nu\geq 6\:\mathrm{eV}$ initially increase in magnitude as much as $10\%$, in contrast to the marked decrease of the global maximum. Yet, with the exception of the sample at $10\: \mathrm{eV}$ subsequent electron exposures regain the decreasing pattern of the maximum, albeit to a lesser magnitude, and rate of change (see figure \ref{fig.e1-uv-diff}). After relaxation, however, some degree of elasticity is observed, with  most of the test samples recovering to the initial state, although exposing the relaxed material anew to electron irradiation results in an increment of the permittivity value in clear contrast to the global maximum bearing.\\    

Concerning the imaginary part of the permittivity, the previous description gets somewhat inverted when compared to the measurements of $\e_1$ in both ranges. In \emph{range A} the initial behaviour is mixed, with some of the samples increasing in magnitude, while others decrease. Notably, test samples  below the band-gap energy of the material show a similar behaviour to that of the imaginary permittivity in \emph{range A}, an initial increment followed by a steady  decrease with new dosages. \emph{Range B} on the other hand shows a behaviour close to the comportment of the global maximum, with the anomaly of test sample at $10\; \mathrm{eV}$. After relaxation the degree of recovery is mixed, in \emph{range B} the hysteretic behaviour is conspicuous.  But in \emph{range A} the results are  to some extent extraordinary,   all the samples beyond the band-gap of the material show some degree of  recovery,  in some cases surpassing the initial value by as much as $15\%$ for example at $3.6\:\mathrm{eV}$; below the band-gap energy test samples show an eerie result not only there is no recovery to the initial state, but a considerable increment instead.  After relaxation all samples show the same decreasing pattern as observed in their real counterpart.  In \emph{range B} all the samples follow a similar pattern to that of the maximum; an initial fast decrease, followed by a sustained decrease in magnitude, although steadily reducing the rate of change. The physical explanation to this peculiar reaction of the material, reflected by the permittivity, in response to electron bombardment can be reasoned in term of the modification to the physical bonding between the Arsenic and Sulfide atoms, and the electron traps within the material, assuming that there is no increment in the initial number of  electrons in the material, which to this effect is grounded. \\

\subsection{Physical process\label{sec.explanation}}

\begin{figure}
\includegraphics[width=0.7\textwidth]{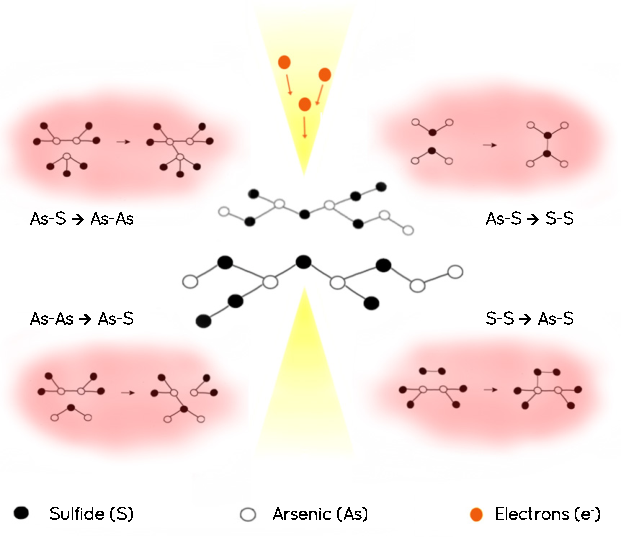}
\caption{Homopolar and heteropolar bond breaking process under electron irradiation in \film . \label{fig.process}}
\end{figure}

To fathom the physical process behind this phenomenon it is necessary to gain some understanding about the atomic configuration of \film~ chalcogenide glass. Either amorphous or crystalline, the structure of chalcogenide material based on As and S is composed of pyramidal arrangement in threads that form layers linked by homopolar (As--As and S--S) and heteropolar (As--S) bonds (figure \ref{fig.process}). The bonding energies in the homopolar case are $1.88\:\mathrm{eV}$ and $2.39\:\mathrm{eV}$ for As--As and S--S bonds, respectively; whilst for the heteropolar (As--S) the bonding energy is $2.29\:\mathrm{eV}$ \cite{Ramirez-Malo1994}. Our film has  similar proportions of As and S atoms, $51\%$ and $49\%$ respectively, as measured by XPS in the surface of the material. To evaluate the overall bulk composition of the film we use EDX finding atomic proportions closer to that of \film~chalcogenide glass  (As $58\%$ and S $42\%$). The atomic composition measured before and after the experiment by EDX did not show any net flux of atom concentration.  It is important to note that our  TEM instrument has the ability to do STEM, TEM and EDX within the same chamber, thus avoiding the possibility of contamination or oxidation of the sample by keeping it under high vacuum. Under irradiation exposure the incoming electrons ($300\:\mathrm{keV}$) can break the homopolar and heteropolar bonds with ease by mean of inelastic collisions; whence a structural rearrangement follows, with the broken homopolar bonds switching to heteropolar bonds \cite{Nordman1998,Kovalskiy2008}. This structural rearrangement will continue for as long as the material keeps being bombarded with electrons. However, since the concentration of As atoms is higher, the re-bonding of the S and As atoms, mostly coming from the broken homopolar bonds, will lead to dangling uncoordinated As atoms. Under relaxation, the dangling As atoms, in vacuum, re-bond in homopolar As--As pairs. Consequently, the density of homopolar bonded As  increases, resulting in a diminished response to electron irradiation as observed in the decrement in the rates of change before and after the first irradiation and relaxation period \ref{fig.derivatives}.\\  

Concomitant to the re-bonding process is the nano-crystallization of the material, which causes a reduction in the number of the trapped carriers, either by promoting a large fraction of the electrons to the conduction band, or by reducing the number of carrier traps in the material. The overall  modification of the electronic density states  originates  the prominent reduction in the permittivity. Recall that in amorphous and semi-crystalline materials the permittivity heavily depends on the density of trapped carriers. This comes from the fact that carrier transport in these materials, like chalcogenide glass, is controlled by traps; at any given point in time a fraction of the carriers is confined, and since the dipole moment of the filled and empty traps may vary broadly it changes the permittivity of the material \cite{Andriesh2004,Nastas2006}. A working model for the dielectric constant based on the electron trapping was given by Arkhipov \etal \cite{Arkhipov1981},

\begin{equation}\label{eq.dtrap}
\e(\vec r,t)=\e_0+4\pi\kappa_0\int_0^\infty \rho(\vec r,t;E)dE,
\end{equation}\\
\noindent where $\rho\:dE$ is the carrier density trapped in the energy interval $E$ to $E+dE$, $\vec r$ is the position vector, $t$ the time, $E$ is the trap energy, $\e_0$ is the vacuum permittivity, and $\kappa_0$ is a coefficient that depicts the change in the dipole moment of the traps due to the capturing of electrons in them. Observe that in order to reduce the permittivity at $t>t_0$ a reduction in the density of trapped electrons is inexorable, \ie $\rho(\vec r, t,; E)<\rho(\vec r, t_0,; E)$.\\

Evidence of such electronic rearrangement under energy absorption by \film~chalcogenide thin film has been reported separately by Tanaka \etal~and Lee \etal. Tanaka \etal~\cite{Tanaka1987}, observed chemical and medium range re-ordering in \film~film under photon energy absorption. Meanwhile, using X-rays, Lee \etal~observed modification in the structure order of chalcogenide films \cite{Lee1989}. Albeit focusing on the dynamics of photo-darkening and anisotropy, respectively, both studies show that upon energy absorption the material suffers an alteration in both the electron energy density and the bond structure; in agreement with our observations, where the alteration of the permittivity springs from the changes in the structural and electronic states of the film.\\

\subsection{Prospective deduction, the refractive index}   

\begin{figure}[hb]
\centering
\subfloat[$n(\e_1,\e_2)$]{\includegraphics[width=0.5\textwidth]{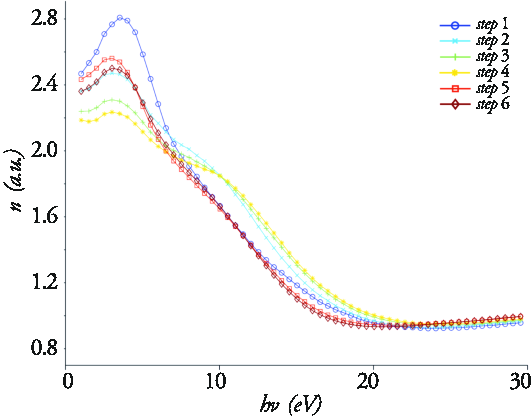}\label{fig.e-n}}
\subfloat[$k(\e_1,\e_2)$]{\includegraphics[width=0.5\textwidth]{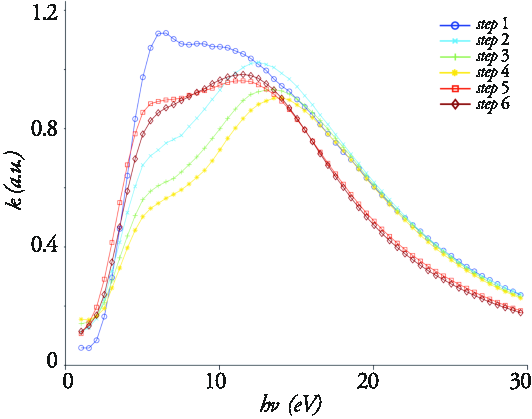}\label{fig.e-k}}
\caption{(a) Refractive index $n$, and (b) absorption constant $k$, derived from the real and imaginary permittivity.\label{fig.e-nk}}
\end{figure}

Earlier we discussed the inextricable relation between the refractive index, $n$, and absorption, $k$, with the electromagnetic properties of the material, $\e_r$ and $\mu_r$. Often, in the deduction of the refractive index, the permeability, $\mu_r$, is set to a constant value, generally $1$, resulting in a set of relations widely available in the basic literature. Here we present the calculated  optical parameters, $n$ and $k$, based on these relations (see figures \ref{fig.e-n} and \ref{fig.e-k} respectively).\\

\begin{figure}[hb]
\subfloat[$n_{max}(\e_1,\e_2)$]{\includegraphics[width=0.5\textwidth]{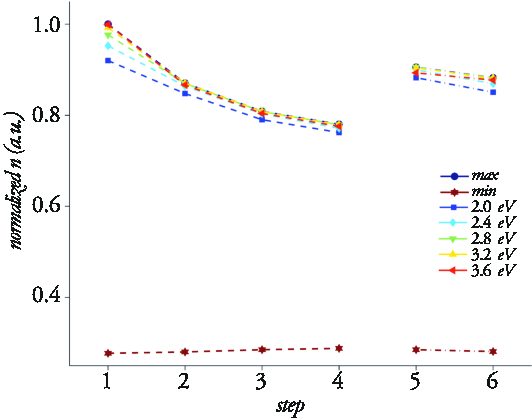}\label{fig.e-nmax}}
\subfloat[$k_{max}(\e_1,\e_2)$]{\includegraphics[width=0.5\textwidth]{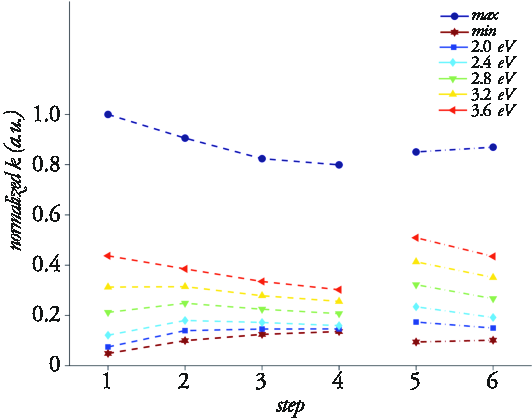}\label{fig.e-kmax}}
\caption{Peak trace sampling of (a) refractive index, $n$, and (b) absorption constant, $k$, derived from the real and imaginary permittivity.\label{fig.e-nkmax}} 
\end{figure}

Under the former assumption, the computed results show similar behaviour for $n$ and $k$ to that observed for $\e_1$ and $\e_2$, with the peak of the refractive index, $n$,  reducing by $\sim 23\%$ after an irradiation time of $t=1s$. The minimum change takes place at $6.5\:\mathrm{eV}$ where the reduction is $\sim 8\%$. For energies close to the band-gap the reduction is on average  $\sim 20\%$. The absorption, $k$, also shows a striking decay, down to $40\%$ from its peak value, and a minimum of $\sim 10\%$, with an average reduction of $\sim 35\%$ for dispersion energies close to the band-gap. Calculation of the extrema of the refractive index and absorption coefficient are shown in figures \ref{fig.e-nmax} and \ref{fig.e-kmax}.\\

The observed measurements, and the calculated optical properties therein, are extraordinary, in the sense that all previously published experiments, with high energy electrons ($40\:\mathrm{keV}$), reported an increase in the refractive index between $3\%$ to $8\%$ \cite{Suhara1975,Nishihara1978,Nordman1996a,Nordman2001a,Nordman2001,Tanaka1997a}. In contrast to the vast literature reporting photon-induced refractive index change, which has yielded as much as an $8\%$ increase in the refractive index of \film~film under illumination \cite{DeNeufville1974, Tanaka2009},  in our experiment the refractive index decreases as much as $23\%$. Bearing in mind that the conditions in the cited experiments are significantly different from those reported here, we believe the discrepancy could be explained by two different mechanisms. The first would state that, as described in the previous section, the permeability remains unchanged, while the number of electrons and the number of homopolar and broken heteropolar bonds increases; leading to the recombination mechanism described earlier. These structural alterations, in turn, induce nano-crystallization, ultimately leading to the generalized reduction in the number of energy traps, especially within  energies in \emph{range A}; all these simultaneous changes will cause  reduction in the permittivity, and therefore reduce the refractive index.\\

On the other hand, to reconcile  previous published results with ours, it would be necessary to acknowledge a dynamic change of the permeability with respect to electron irradiation. This would require the electrons to alter the atoms intrinsically by means of elastic collisions, and/or induce current loops,  causing the magnetic dipoles in the material to reorganize, so inducing paramagnetic states, and hence increasing the permeability.  However plausible the latter explanation, an experimental confirmation is required, together with further studies on the permeability of chalcogenide glass under high energy electron- and photon- irradiation.\\

\section{Conclusions\label{sec.fin}}

A new characterization procedure and analysis of the permittivity of \film~chalcogenide glass has been presented based on low-loss Electron Energy Loss Spectroscopy. The results are extendable to the optical regime by means of  the small angle scattering approximation. Furthermore, they allow us to calculate an approximate form of the refractive index, assuming constant permeability, and suggest the possibility of magnetic alterations induced by electron irradiation.\\%

The calculated results and observations found that high energy electrons induced a reduction in the permittivity, real and imaginary, of \film~thin film. The real permittivity underwent a maximum reduction of $\sim 40\%$, while the imaginary permittivity decreased by $\sim 50\%$. The results can be explained in terms of the atomic bond reconfiguration; in this model the incident electrons break the homopolar and heteropolar bonds, leading to a reduction of the former, \emph{correcting} the \emph{wrong} bonds.\\

The results are significant to the development of manifold photonic applications, with applications to numerous areas of research and engineering. Namely, the observed reduction in the permittivity could enable a new range of transformation optic devices, which have been so far limited to the realm of far-IR range of the electromagnetic field. Furthermore, these results could be significant to future implementation of reconfigurable photonic circuits, infrared telecommunications, photonic crystals, and all optical conversion and computing.\\

\section*{Acknowledgements}

The authors wish to acknowledge Erica T. Jolly , Alfonso Caraveo, and Carlos Argaez for their suggestions, corrections and valuable discussions in preparing this manuscript.


\begin{thebibliography}{10}

\bibitem{Yang2010}
Z.~Yang, M.~K. Fah, K.~A. Reynolds, J.~D. Sexton, M.~R. Riley, M.-L. Anne,
  B.~Bureau, and P.~Lucas, ``{Opto-electrophoretic detection of bio-molecules
  using conducting chalcogenide glass sensors},'' {\em Optics Express},
  vol.~18, p.~26754, Dec. 2010.

\bibitem{Eggleton2011}
B.~J. Eggleton, B.~Luther-Davies, and K.~Richardson, ``{Chalcogenide
  photonics},'' {\em Nature Photonics}, vol.~5, pp.~141--148, 2011.

\bibitem{Pelusi2010a}
M.~D. Pelusi, F.~Luan, S.~J. Madden, D.-y. Choi, D.~Bulla, B.~Luther-Davies,
  and B.~J. Eggleton, ``{Chalcogenide Glass Chip Based Nonlinear Signal
  Processing - OSA Technical Digest (CD)},'' in {\em Integrated Photonics
  Research, Silicon and Nanophotonics}, p.~IWC3, Optical Society of America,
  July 2010.

\bibitem{Eggleton2010}
B.~J. Eggleton, ``{Chalcogenide photonics: fabrication, devices and
  applications Introduction},'' {\em Optics Express}, vol.~18, p.~26632, Dec.
  2010.

\bibitem{Xiong2010}
C.~Xiong, L.~G. Helt, A.~C. Judge, G.~D. Marshall, M.~J. Steel, J.~E. Sipe, and
  B.~J. Eggleton, ``{Quantum-correlated photon pair generation
in
  chalcogenide As\_2S\_3 waveguides},'' {\em Optics Express}, vol.~18,
  p.~16206, July 2010.

\bibitem{Bendana2011}
X.~Benda\~{n}a, A.~Polman, and F.~J. {Garc\'{\i}a de Abajo}, ``{Single-photon
  generation by electron beams.},'' {\em Nano letters}, vol.~11, pp.~5099--103,
  Dec. 2011.

\bibitem{Juodkazis2004}
S.~Juodkazis, T.~Kondo, and H.~Misawa, ``{Three-dimensional recording and
  structuring of chalcogenide glasses by femtosecond pulses},'' in {\em
  Proceedings of SPIE}, vol.~5662, pp.~179--184, SPIE, Oct. 2004.

\bibitem{Chen1970}
G.~Chen, H.~Jain, M.~Vlcek, and A.~Ganjoo, ``{Photoinduced volume change in
  arsenic chalcogenides by bandgap light},'' {\em Phys. Rev. B}, vol.~74,
  p.~174203, 2006.

\bibitem{Kozak2009a}
M.~I. Kozak, V.~Y. Loya, N.~P. Golub, and M.~Y.
  Onis\^{a}$\backslash$euro$\backslash$texttrademarkko, ``{Mechanism of
  photoinduced nanodimensional expansion/contraction in glassy thin layers of
  As2S3},'' {\em Theoretical and Experimental Chemistry}, vol.~45, pp.~69--73,
  May 2009.

\bibitem{Fritzsche1996a}
H.~Fritzsche and H.~Firtzsche, ``{Photo-induced fluidity of chalcogenide
  glasses},'' {\em Solid State Communications}, vol.~99, no.~3, p.~153, 1996.

\bibitem{Tanaka2005}
K.~Tanaka, ``{Photoinduced deformations in chalcogenide glasses: scalar and
  vectorial},'' {\em Journal of Optoelectronics and Advanced Materials},
  vol.~7, no.~5, pp.~2571 -- 2580, 2005.

\bibitem{Feinleib1971}
J.~Feinleib, J.~P. DeNeufville, S.~C. Moss, and S.~R. Ovshinsky, ``{Rapid
  reversible light-induced crystallization of amorphous semiconductors},'' {\em
  Applied Physics Letters}, vol.~18, p.~254, Oct. 1971.

\bibitem{Tanaka2009}
K.~Tanaka and K.~Shimakawa, ``{Chalcogenide glasses in Japan: A review on
  photoinduced phenomena},'' {\em physica status solidi (b)}, vol.~246,
  pp.~1744--1757, Aug. 2009.

\bibitem{Istvan2007}
I.~Istvan, {\em {Photo- and ion-induced changes in amorphous chalcogenide
  films}}.
\newblock PhD thesis, Debrecen, 2007.

\bibitem{DeNeufville1974}
J.~{De Neufville}, S.~Moss, and S.~Ovshinsky, ``{Photostructural
  transformations in amorphous As2Se3 and As2S3 films},'' {\em Journal of
  Non-Crystalline Solids}, vol.~13, no.~2, pp.~191--223, 1974.

\bibitem{Gopal1982}
V.~Gopal, ``{Energy gap-refractive index interrelation},'' {\em Infrared
  Physics}, vol.~22, no.~5, pp.~255--257, 1982.

\bibitem{Kurtz2009}
R.~M. Kurtz, W.~Lu, J.~Piranian, T.~Jannson, and A.~O. Okorogu, ``{The Fast
  Photorefractive Effect and Its Application to Vibrometry},'' {\em Journal of
  Holography and Speckle}, vol.~5, pp.~149--155, Aug. 2009.

\bibitem{VKTikhomirovandSRElliott1995}
{V K Tikhomirov and S R Elliott}, ``{The anisotropic photorefractive effect in
  bulk As 2 S 3 glass induced by polarized subgap laser light},'' {\em Journal
  of Physics: Condensed Matter}, vol.~7, no.~8, p.~1737, 1995.

\bibitem{Submitted2010}
M.~Kowalyshen, {\em {Photoinduced Dichroism in Amorphous As2Se3 Thin Film}}.
\newblock PhD thesis, 2010.

\bibitem{Ta'eed2006}
V.~G. Ta'eed, M.~R.~E. Lamont, D.~J. Moss, B.~J. Eggleton, D.-Y. Choi,
  S.~Madden, and B.~Luther-Davies, ``{All optical wavelength conversion via
  cross phase modulation in chalcogenide glass rib waveguides.},'' {\em Optics
  express}, vol.~14, pp.~11242--7, Nov. 2006.

\bibitem{Lyubin2006}
V.~Lyubin, M.~Klebanov, M.~Veinger, I.~Lyubina, and B.~Sfez,
  ``{Photoluminescence and photostructural transformations in neodymium-doped
  glassy chalcogenide films},'' {\em Optical Materials}, vol.~28,
  pp.~1115--1117, June 2006.

\bibitem{Anderson1975}
P.~Anderson, ``{Model for the electronic structure of amorphous
  semiconductors},'' {\em Physical Review Letters}, vol.~34, no.~15,
  pp.~953--955, 1975.

\bibitem{Fritzsche1998}
H.~Fritzsche, ``{Toward understanding the photoinduced changes in chalcogenide
  glasses},'' {\em Semiconductors}, vol.~32, pp.~850--854, Aug. 1998.

\bibitem{Simdyankin2004}
S.~Simdyankin, S.~Elliott, Z.~Hajnal, T.~Niehaus, and T.~Frauenheim,
  ``{Simulation of physical properties of the chalcogenide glass As2S3 using a
  density-functional-based tight-binding method},'' {\em Physical Review B},
  vol.~69, Apr. 2004.

\bibitem{Andriesh2004}
A.~Andriesh, M.~Iovu, and S.~Shutov, {\em {Semiconducting Chalcogenide Glass II
  - Properties of Chalcogenide Glasses}}, vol.~79 of {\em Semiconductors and
  Semimetals}.
\newblock Elsevier, 1st~ed., 2004.

\bibitem{Singh2007}
J.~Singh and K.~Tanaka, ``{Photo-structural changes in chalcogenide glasses
  during illumination},'' {\em Journal of Materials Science: Materials in
  Electronics}, vol.~18, pp.~423--428, Mar. 2007.

\bibitem{Suhara1975}
T.~Suhara, H.~Nishihara, and J.~Koyama, ``{Electron-Beam-Induced
  Refractive-Index Change of Amorphous Semiconductors},'' {\em Japanese Journal
  of Applied Physics}, vol.~14, pp.~1079--1080, July 1975.

\bibitem{Nordman1996a}
N.~Nordman and O.~Salminen, ``{Thickness variations in amorphous As2S3 films
  induced by electron beam},'' {\em Solid State Communications}, vol.~100,
  pp.~241--244, Oct. 1996.

\bibitem{Nordman2001a}
N.~Nordman and O.~Nordman, ``{Refractive index change caused by electron
  irradiation in amorphous As–S and As–Se thin films coated with different
  metals},'' {\em Journal of Applied Physics}, vol.~90, p.~2206, Sept. 2001.

\bibitem{Nordman2001}
O.~Nordman, N.~Nordman, and V.~Pashkevich, ``{Refractive-index change caused by
  electrons in amorphous AsS and AsSe thin films doped with different metals by
  photodiffusion},'' {\em Journal of the Optical Society of America B},
  vol.~18, p.~1206, Aug. 2001.

\bibitem{Tanaka1997a}
K.~Tanaka, ``{Electron beam induced reliefs in chalcogenide glasses},'' {\em
  Applied Physics Letters}, vol.~70, p.~261, Jan. 1997.

\bibitem{Perrin1974}
J.~Perrin, J.~Cazaux, and P.~Soukiassian, ``{Optical Constants and Electronic
  Structure of Crystalline and Amorphous As2S3 in the 3 to 35 eV Range},'' {\em
  physica status solidi (b)}, vol.~62, pp.~343--350, Apr. 1974.

\bibitem{Egerton1996}
R.~F. Egerton, {\em {Electron Energy Loss Spectroscopy in the Electron
  Microscope}}.
\newblock New York: Plenum Press, 2nd~ed., 1996.

\bibitem{Ritchie1957}
R.~H. Ritchie, ``{Plasmon losses by fast electron in thin films},'' {\em
  Physical Review}, vol.~106, pp.~874--881, 1957.

\bibitem{Nozieres1959}
P.~Nozieres and D.~Pines, ``{Electron interaction in solids: Chacteristic
  energy-loss spectrum},'' {\em Physical Review}, vol.~113, pp.~1254--1267,
  1959.

\bibitem{Stoger-Pollach2008}
M.~St\"{o}ger-Pollach, ``{Optical properties and bandgaps from low loss EELS:
  pitfalls and solutions.},'' {\em Micron (Oxford, England : 1993)}, vol.~39,
  pp.~1092--110, Dec. 2008.

\bibitem{Egerton2009}
R.~F. Egerton, ``{Electron energy-loss spectroscopy in the TEM},'' {\em Reports
  on Progress in Physics}, vol.~72, p.~016502, Jan. 2009.

\bibitem{Verbeeck2009}
J.~Verbeeck and G.~Bertoni, ``{Deconvolution of core electron energy loss
  spectra.},'' {\em Ultramicroscopy}, vol.~109, pp.~1343--52, Oct. 2009.

\bibitem{GarciadeAbajo2010}
F.~J. {Garc\'{\i}a de Abajo}, ``{Optical excitations in electron microscopy},''
  {\em Reviews of Modern Physics}, vol.~82, pp.~209--275, Feb. 2010.

\bibitem{Zhang2010b}
L.~Zhang, S.~Turner, and J.~Verbeeck, ``{Model-based determination of
  dielectric function by STEM low-loss EELS},'' {\em Physical Review B},
  vol.~81, p.~035102, Jan. 2010.

\bibitem{Schafer2012}
K.~Hoffmann, {\em {Electron Energy Loss Spectroscopy as an Experimental Probe
  for the Crystal Structure and Electronic Situation of Solids.}}
\newblock Weinheim, Germany: Wiley-VCH Verlag GmbH \& Co. KGaA, Apr. 2012.

\bibitem{Ramirez-Malo1994}
J.~Ram\'{\i}rez-Malo, E.~M\'{a}rquez, C.~Corrales, P.~Villares, and
  R.~Jim\'{e}nez-Garay, ``{Optical characterization of As2S3 and As2Se3
  semiconducting glass films of non-uniform thickness from transmission
  measurements},'' {\em Materials Science and Engineering: B}, vol.~25,
  pp.~53--59, June 1994.

\bibitem{Nordman1998}
O.~Nordman, N.~Nordman, and N.~Peyghambarian, ``{Electron beam induced changes
  in the refractive index and film thickness of amorphous As[sub x]S[sub
  100−x] and As[sub x]Se[sub 100−x] films},'' {\em Journal of Applied
  Physics}, vol.~84, p.~6055, Dec. 1998.

\bibitem{Kovalskiy2008}
A.~Kovalskiy, J.~Neilson, A.~Miller, F.~Miller, M.~Vlcek, and H.~Jain,
  ``{Comparative study of electron- and photo-induced structural
  transformations on the surface of As35S65 amorphous thin films},'' {\em Thin
  Solid Films}, vol.~516, pp.~7511--7518, Sept. 2008.

\bibitem{Nastas2006}
a.~M. Nastas, a.~M. Andriesh, V.~V. Bivol, a.~M. Prisakar, and G.~M. Tridukh,
  ``{Effect of electric field on photoinduced changes in the optical properties
  of chalcogenide glassy semiconductors},'' {\em Technical Physics Letters},
  vol.~32, pp.~45--47, Jan. 2006.

\bibitem{Arkhipov1981}
V.~Arkhipov, M.~Iovu, M.~Iovu, A.~Rudenko, and S.~Shutov, ``{Negative transient
  currents in amorphous semiconductors},'' {\em International Journal of
  Electronics}, vol.~51, pp.~735--742, 1981.

\bibitem{Tanaka1987}
K.~Tanaka, ``{Chemical and medium-range orders in As2S3 glass},'' {\em Physical
  Review B}, vol.~36, no.~18, pp.~9746--9752, 1987.

\bibitem{Lee1989}
J.~M. Lee, G.~Pfeiffer, M.~A. Paesler, D.~E. Sayers, and A.~Fontaine, ``{Photon
  intensity-dependent darkening kinetics in optical and structural anisotropy
  in a-As2S3: A study of X-ray absorption spectroscopy},'' {\em Journal of
  Non-Crystalline Solids}, vol.~114, pp.~52--54, 1989.

\bibitem{Nishihara1978}
H.~Nishihara, Y.~Handa, T.~Suhara, and J.~Koyama, ``{Direct writing of optical
  gratings using a scanning electron microscope},'' {\em Applied Optics},
  vol.~17, p.~2342, Aug. 1978.

\end{thebibliography}
\end{document}